\documentclass[onecolumn,journal]{IEEEtran}

\usepackage[T1]{fontenc}
\usepackage{lmodern}
\usepackage{cite}
\usepackage{amsmath,amssymb}
\usepackage{graphicx}
\usepackage{url}
\usepackage{booktabs}
\usepackage{orcidlink}

\usepackage{xcolor}
\usepackage{listings}
\usepackage{fancyvrb}

\definecolor{codebg}{RGB}{246,248,250}
\definecolor{codeframe}{RGB}{208,215,222}
\definecolor{codeborder}{RGB}{208,215,222}
\definecolor{codetext}{RGB}{36,41,47}
\definecolor{codekeyword}{RGB}{207,34,46}
\definecolor{codetype}{RGB}{149,56,0}
\definecolor{codestring}{RGB}{14,98,69}
\definecolor{codecomment}{RGB}{106,115,125}
\definecolor{codelabel}{RGB}{87,96,106}

\lstdefinestyle{appendix}{
  backgroundcolor=\color{codebg},
  basicstyle=\small\ttfamily\color{codetext},
  breaklines=true,
  breakatwhitespace=false,
  columns=fullflexible,
  keepspaces=true,
  showstringspaces=false,
  frame=single,
  rulecolor=\color{codeborder},
  framerule=0.4pt,
  framesep=6pt,
  xleftmargin=0pt,
  xrightmargin=0pt,
  aboveskip=0.6em,
  belowskip=0.8em,
  tabsize=2,
  keywordstyle=\color{codekeyword}\bfseries,
  stringstyle=\color{codestring},
  commentstyle=\color{codecomment}\itshape,
  emphstyle=\color{codetype}\bfseries,
}

\lstdefinelanguage{yaml}{
  sensitive=false,
  comment=[l]{\#},
  morestring=[b]",
  morestring=[b]',
  literate=
    {:}{{\color{codekeyword}:}}{1}
    {-}{-}{1}
}

\lstdefinelanguage{BAML}{
  keywords={enum,class,function,client,prompt,string,test,functions,args,true,false,this},
  keywordstyle=\color{codekeyword}\bfseries,
  sensitive=true,
  comment=[l]{//},
  morestring=[b]",
  morecomment=[s]{@"}{"},
  emph={Severity,Summary,SummarizeTicket,TicketModel,P1,P2,P3,description},
  emphstyle=\color{codetype},
}

\newcommand{\codelabel}[1]{%
  \vspace{0.4em}%
  {\small\sffamily\color{codelabel}#1\par\vspace{0.15em}}%
}

\newcommand{\codeinput}[3]{%
  \codelabel{#1}%
  \lstinputlisting[style=appendix, language=#3]{#2}%
}

\usepackage{tikz}
\usetikzlibrary{arrows.meta,positioning,fit,calc}

\definecolor{specfill}{RGB}{225,245,238}
\definecolor{specborder}{RGB}{15,110,86}
\definecolor{agentfill}{RGB}{238,237,254}
\definecolor{agentborder}{RGB}{83,74,183}
\definecolor{artifactfill}{RGB}{250,236,231}
\definecolor{artifactborder}{RGB}{153,60,29}
\definecolor{forbiddenred}{RGB}{163,45,45}

\begin{document}

\title{{\fontsize{18}{22}\selectfont From Prompt Engineering to Programmatic Prompting:\\
Toward a Software Engineering Discipline for Large Language Models}}

\author{Sebastian~Dingler\,\orcidlink{0000-0002-0162-8428}\\ Stuttgart, Germany%
\thanks{For correspondence, remove \texttt{foobar} from \texttt{s.dinglerfoobar@gmail.com}.}}

\maketitle

\begin{abstract}
The practice of instructing large language models (LLMs) through natural-language prompts has, in the span of a few years, migrated from an interactive curiosity to a load-bearing component of production software. Yet the vocabulary, tooling, and mental models of the field remain anchored in its interactive origins: we still speak of ``prompt engineering'' as though the crafting of a single well-worded instruction were the central problem. This paper argues for a categorical distinction. Prompt engineering, we contend, is an \emph{interaction technique}---a human-computer interaction skill for eliciting desired behavior from a model in a conversational or exploratory setting. Programmatic prompting, by contrast, is an \emph{engineering discipline} governing the behavioral interface between software systems and language models: the construction, composition, versioning, testing, and maintenance of prompts as first-class software artifacts embedded in larger systems. Conflating the two has produced brittle applications, unrepeatable evaluations, and an accumulation of \emph{prompt debt}. We make three contributions: (1)~a definition of programmatic prompting; (2)~a systematic differentiation of programmatic prompting from prompt engineering, prompt tuning, fine-tuning, and general LLM application development; and (3)~an account of prompt debt that distinguishes the self-admitted debt measurable in code from the silent debt that surfaces only as incidents, and identifies its economic cause. Our thesis is simple to state and consequential to accept: prompt engineering is an interaction technique; programmatic prompting is an engineering discipline.
\end{abstract}


\section{Introduction}
\IEEEPARstart{E}{very} computing abstraction that has mattered has passed through a phase in which its practitioners mistook interaction for engineering. Early programming was ``coding'' in the literal sense---the manual translation of algorithms into machine instructions---until the software crisis of the late 1960s forced the community to recognize that constructing large, long-lived programs was a fundamentally different activity than writing individual routines, and the discipline of software engineering was born~\cite{ref1}. That history marks a fork every interactive practice reaches when it outgrows its origins: it either develops an engineering discipline or pays permanently for the absence of one.

We argue that prompting large language models is at precisely this juncture. The term \emph{prompt engineering} entered common use around 2021--2022 to describe the craft of phrasing inputs so that a model produces desired outputs~\cite{ref3,ref4}. Its rise had an economic logic: Brown et al.\ had shown that in-context instruction could approach the performance of task-specific fine-tuning at zero training cost~\cite{ref7}, making the prompt the cheapest route to specialized model behavior---adjustable at inference time, by anyone, without a training pipeline. The term was apt for its moment: practitioners were exploring what models could do, and the unit of work was the individual prompt, iterated by hand in a playground or chat interface. But the moment has passed. Prompts today are embedded in retrieval pipelines, agent loops, multi-model routing systems, and customer-facing products. They are concatenated from templates, populated with untrusted user input, chained across dozens of model calls, and expected to behave consistently across model versions the developer does not control. The individual prompt is no longer the unit of work; the \emph{prompt-based system} is.

The vocabulary has not kept pace, and the consequences are not merely terminological. When an organization believes its LLM problems are prompt-engineering problems---problems of wording---it staffs, tools, and budgets accordingly: a clever writer, a playground, an afternoon of iteration. When the problems are in fact engineering problems---problems of specification, composition, regression, and drift---this response fails in characteristic ways, and the failures are now in the public record. Prompts are edited in production without version control: in May 2025, one unreviewed change to the system prompt of xAI's Grok bot pushed politically charged content into thousands of unrelated conversations, and the announced remedy---published prompts and mandatory review of prompt changes---was, precisely, version control~\cite{ref24}. A wording change that fixes one behavior silently breaks three others, and no test suite exists to notice; an unevaluated support chatbot invents a refund policy, and a tribunal holds the company to it, as Air Canada learned in the first such ruling~\cite{ref25}. A model provider ships an update and an application's behavior shifts overnight, with no baseline against which to measure the shift---a drift measured in the field when GPT-4's accuracy on one fixed task fell from 84\% to 51\% between its March and June 2023 versions, both versions meanwhile acquiring formatting habits that broke previously executable code output~\cite{ref26}. Evaluation, where it exists at all, consists of a developer eyeballing a handful of outputs---a practice documented among non-experts by Zamfirescu-Pereira et al.~\cite{ref5}, confirmed among professional developers by Liang et al.~\cite{ref22}, and shown by Shankar et al.\ to be unreliable even when done carefully~\cite{ref6}.

This paper's thesis is that the field is already implicitly developing two distinct activities, and that their separation---still blurred in practice under shared vocabulary---should now be made explicit:
\begin{quote}
\textbf{Prompt engineering is an interaction technique. Programmatic prompting is an engineering discipline.}
\end{quote}

The first is a skill of human-computer interaction: knowing how to phrase, decompose, and iterate on instructions to elicit desired behavior from a model, in real time, with a human in the loop to judge the result. It is genuinely valuable, genuinely learnable, and genuinely not engineering---in the same way that skilled use of a search engine, a calculator, or a REPL (read--eval--print loop, the interactive console in which programmers try out code line by line) is valuable to an engineer without being engineering itself. The second is the discipline of treating prompts as software: artifacts with specifications, interfaces, tests, versions, dependencies, performance characteristics, and failure modes, embedded in systems that must operate without a human judging every output. Its relation to the operational tooling now marketed as \emph{LLMOps} is that of software engineering to DevOps---complementary, never substitutable: dashboards and cost tracking can show that a monitored system's behavior changed, but without specifications and evaluation baselines they cannot say whether it is correct.

The remainder of this paper develops this argument. Section~\ref{sec:background} reviews the current landscape and its terminological confusion. Section~\ref{sec:definition} defines programmatic prompting. Section~\ref{sec:not} systematically distinguishes it from its neighbors. Section~\ref{sec:related} situates the argument among related work, and Section~\ref{sec:conclusion} concludes with the research problems the discipline now faces.

\section{Background: A Field Outgrowing Its Vocabulary}
\label{sec:background}

\subsection{The Interactive Origins of Prompting}
In-context learning---the observation that large models can perform novel tasks when given instructions and examples in their input, without weight updates---was established as a paradigm by Brown et al.~\cite{ref7}. The immediate practical consequence was that model behavior became \emph{programmable through natural language at inference time}, and a body of technique rapidly accumulated: few-shot exemplar selection, chain-of-thought elicitation~\cite{ref8}, role assignment, output-format scaffolding, and a growing catalog of reusable patterns~\cite{ref4}. Reynolds and McDonell presciently framed prompting as ``programming in natural language'' as early as 2021~\cite{ref3}, but the dominant practice remained interactive: a human writes a prompt, reads the output, and revises.

This interactive practice is what most people mean by prompt engineering, and it has the epistemic structure of interaction, not construction. The feedback loop is seconds long. The evaluator is a human reading a single output. Success is local: \emph{this} prompt worked on \emph{this} input against \emph{this} model \emph{today}. Nothing in the practice ensures---or even asks whether---the prompt works on inputs the author did not try, composes with other prompts, survives a model update, or transfers across different model vendors. In machine-learning terms, interactive prompting is manual overfitting: the author optimizes the prompt against the visible handful of inputs---the training set---with no held-out validation, so apparent success measures fit to those examples, not performance on the task. Yet for any recurring task, the property that actually matters is generalization to the unseen input distribution, and generalization is precisely what a validation-free process cannot measure. The golden datasets that the discipline will mandate (Section~\ref{sec:definition}) are, in this framing, nothing more exotic than the missing validation sets---and their absence is not a small validation set but the lack of any defined quantity to estimate, a distinction Section~\ref{sec:versus-pe} returns to.

\subsection{The Shift to Embedded Prompts}
Contemporary LLM applications embed prompts inside software. A customer-support agent may involve a routing prompt, a retrieval-augmented answering prompt, a tool-selection prompt, a safety-review prompt, and a summarization prompt, executed in a graph with branching and retries. Frameworks such as LangChain popularized such chains; research systems such as AI Chains~\cite{ref9} and PromptChainer~\cite{ref10} studied their authoring; and languages and compilers such as LMQL~\cite{ref11} and DSPy~\cite{ref12} have begun to treat prompt programs as compilable artifacts whose concrete wording is an optimization target rather than a hand-crafted product.

In these systems, the properties that matter are systemic: end-to-end task success rates, tail-latency and cost budgets, robustness to adversarial or malformed input, stability across model versions, and graceful degradation. None of these properties is visible at the level of an individual prompt, and none can be achieved by wordsmithing alone. Yet the industry's job titles, courses, and tooling still center the individual prompt---a category error this paper aims to name and repair.

\subsection{Prompt Debt}
We use the term \textbf{prompt debt} for the accumulated liability of prompt-based systems built with interactive methods: prompts without owners, tests, or documentation; magic phrasings whose purpose no one remembers (``do not apologize''; ``think step by step''; a stray ``IMPORTANT!!!'' of unknown provenance); implicit couplings between prompts and specific model versions; and evaluation baselines that exist only in a departed employee's memory. Prompt debt is a species of the hidden technical debt Sculley et al.\ diagnosed in machine-learning systems~\cite{ref13}, but with an aggravating property: because prompts are natural language, they \emph{look} self-documenting while being anything but. The incidents catalogued in Section~I are this debt falling due---the unreviewed prompt change, the unevaluated support bot, and the unmeasured model migration each began as an item on the list above. Empirical work has begun to measure the fraction of such debt that developers themselves notice: mining self-admitted technical debt from over 93{,}000 LLM-integration files, Aljohani and Do find prompt-related debt the largest LLM-specific category~\cite{ref27}. But self-admission has a blind spot our argument predicts: where no specification exists, failure has no definition, and where no evaluation exists, regression emits no signal---a developer cannot admit a debt that nothing in the process makes visible. Most prompt debt is therefore \emph{silent} rather than admitted, surfacing not in TODO comments but in production. It also has a structural cause that conventional software engineering never faced: the practice that would make it visible bills per use. A code test suite runs in continuous integration for free, which is why testing culture could become universal; a prompt evaluation suite costs money and time on every run---evaluation-set size, times samples per input under stochastic decoding, times calls per chain---while interactive iteration stays near-free with feedback in seconds. A visible, recurring bill set against an invisible benefit is a textbook recipe for underinvestment, and the cost is shifted rather than avoided: it is repaid as silent debt, at a worse exchange rate. The remedy for technical debt has always been discipline, and discipline requires first naming the activity to be disciplined.

\section{Defining Programmatic Prompting}
\label{sec:definition}

\textbf{Definition.} \textit{Programmatic prompting} is the engineering discipline that governs the behavioral interface between software systems and language models---how intent is specified, rendered into model inputs, composed across model invocations, evaluated, versioned, operated, and evolved---treating the prompts that realize this interface as software artifacts whose correctness, robustness, cost, and maintainability are engineered properties rather than emergent ones.

Governing this interface rests on five load-bearing components, which we elaborate in turn.

\textbf{(1) Prompts as artifacts, not utterances.} In programmatic prompting, a prompt is not a string typed into a text box; it is a source artifact---typically a parameterized template plus associated metadata---that lives in version control, has an owner, carries a change history, and is deployed through the same controlled paths as code. The artifact boundary includes everything that determines model input: system text, templates, exemplar sets, retrieval and formatting logic, decoding parameters, and the identity of the target model. A change to any of these is a change to the artifact.

\textbf{(2) Explicit specification.} Every prompt artifact has a stated contract: the task it performs, the schema and invariants of its inputs, the schema and invariants of its outputs, its tolerated failure modes, and its non-functional budget (latency, token cost). The specification is what makes testing meaningful and what makes the inevitable rewording of a prompt a \emph{refactoring} (behavior-preserving) rather than an unexamined mutation. Because LLM behavior is stochastic and specifications of natural-language tasks are inherently partial, these contracts are typically probabilistic and example-anchored---``$\geq$95\% of outputs on the reference distribution parse as the target schema; factuality score $\geq X$ on the golden set''---rather than universally quantified. This weakens, but does not eliminate, the analogy to conventional interface contracts; managing partial, statistical specifications is itself part of the discipline.

\textbf{(3) Composition with interfaces.} Prompt-based systems are built from multiple model invocations connected by dataflow. Programmatic prompting treats each invocation as a component with a typed interface, so that chains, routers, and agent loops are \emph{architectures} subject to design reasoning---coupling, cohesion, error propagation, idempotency of retries---rather than ad-hoc scripts. Languages such as LMQL~\cite{ref11} and frameworks such as DSPy~\cite{ref12} are early realizations: DSPy in particular separates the \emph{signature} (what a step must do) from the \emph{prompt} (how the model is currently induced to do it), making the wording a compiled, replaceable implementation detail---exactly the interface/implementation split at the heart of software engineering. Industrial systems such as BAML~\cite{ref21}, which compiles prompts declared as typed functions into native clients, push the same idea into production practice.

Interfaces raise a design question the discipline must own: the \emph{locus of contract enforcement}. An output contract can be enforced at decode time, by constraining generation to the schema~\cite{ref11}, or at parse time, by letting the model generate freely and tolerantly repairing the output against the schema, as in BAML's schema-aligned parsing~\cite{ref21}. The choice is a genuine trade-off: decode-time constraints guarantee syntax but can degrade reasoning quality, while parse-time repair preserves free generation at the price of post-hoc coercion---and, by requiring nothing of the provider, decouples structured outputs from vendor-specific function-calling APIs, directly mitigating the cross-vendor portability problem noted in Section~\ref{sec:background}.

\textbf{(4) Empirical, automated evaluation.} Because prompt behavior cannot be derived from the artifact by inspection, evaluation is not a phase but a permanent fixture: golden datasets, scenario suites, property checks (schema validity, refusal correctness, injection resistance), regression gates on every prompt or model change, and calibrated use of LLM-as-judge methods with known validation of the judges themselves~\cite{ref6,ref14}. The unit of evaluation is the distribution of behavior over an input population---never the anecdote.

\textbf{(5) Operation and evolution.} Deployed prompts face drift from three directions: the input distribution shifts, the underlying model shifts, and the surrounding system shifts. Programmatic prompting therefore includes observability (tracing every model call with its full rendered input and output), monitoring of quality proxies in production, staged rollout and rollback of prompt versions, and planned migration procedures for model upgrades. A prompt that cannot be rolled back is not an engineered artifact; it is a live wire.

The components compound. Once a specification (2) and evaluation assets (4) exist, they jointly constitute an objective function---and the wording of a prompt becomes machine-optimizable. \emph{Automatic prompt optimization}~\cite{ref12,ref17}, in which machines search or generate wording that meets the specification, thereby belongs to the discipline's core practice twice over: within evaluation-driven construction, and within evolution (5), where every substrate change invites re-optimization of wording against the new model. The relationship runs in both directions: the optimizers presuppose these assets, and in turn drive their creation, since a team cannot use one without first writing a specification and building an evaluation set. Section~\ref{sec:versus-pt} disentangles this technique from the similarly named but categorically different prompt tuning.

Two clarifications guard against misreading. First, programmatic prompting is not defined by the \emph{use of code}---a prompt embedded as a string literal in application code is not thereby engineered---but by the presence of specification, evaluation, and controlled evolution. Second, it is not a claim that wording ceases to matter. Wording matters enormously; the discipline's claim is that wording must be \emph{governed}---specified, tested, versioned---rather than merely crafted.

\section{What Programmatic Prompting Is Not}
\label{sec:not}

Precision about a new term requires drawing its borders against four neighbors with which it is routinely conflated---or against which it must be weighed.

\subsection{Versus Prompt Engineering}
\label{sec:versus-pe}

Prompt engineering is the interaction technique of eliciting desired model behavior through the design of individual inputs, with a human in the loop evaluating outputs. Field usage is messier than this definition---the term often stretches over templates, evaluation suites, and versioned deployments, whose practitioners are, in our vocabulary, doing programmatic prompting under an older name. The boundary is epistemic rather than tooling-based: a parameterized template arrived at by trial and error and validated on a handful of tried examples remains the interaction technique (Appendix~\ref{app:contrast} gives the contrast in executable form). The distinction is not one of skill level or seriousness but of \emph{category}, and it can be drawn along six axes:

\begin{table}[!t]
\caption{Prompt Engineering Versus Programmatic Prompting}
\label{tab:contrast}
\centering
\renewcommand{\arraystretch}{1.1}
\begin{tabular}{@{}p{0.16\textwidth}p{0.40\textwidth}p{0.40\textwidth}@{}}
\toprule
\textbf{Axis} & \textbf{Prompt engineering} & \textbf{Programmatic prompting} \\
\midrule
Unit of work & An utterance (one prompt, one context) & An artifact and a system of artifacts \\
Feedback loop & Closed by a human reading one output & Closed by evaluation over an input distribution \\
Correctness judged by & The author, per output & Specifications, golden and synthetic datasets, test suites, monitors \\
Success criterion & ``It worked on my input'' & Measured performance on a reference distribution \\
Lifespan of the product & The session & Years, across model generations \\
Failure cost & A retry & An incident \\
\bottomrule
\end{tabular}
\end{table}

The axes are deliberately not independent: they are correlated manifestations of one categorical difference---whether a human judgment or an engineered validation process stands between the prompt and its consequences---and they contrast the two practices as ideal types, not every individual practitioner. Nor is the category a compressed continuum. Coverage is continuous---a reference set of five examples estimates the same quantity as one of five thousand, only worse---but the existence of the quantity is a step function: without a persisted specification, ``regression'' is not poorly measured, it is undefined. Adding examples improves an estimate only where a fixed criterion exists to add them to, and interactive judgment does not converge toward one, however long it runs~\cite{ref22}. A prompt validated against five examples practices the discipline poorly; a prompt validated against none does not practice it at all---and doing something badly differs in kind, not degree, from not doing it.

The relationship between the two mirrors the relationship between exploratory REPL use and software construction. Every good software engineer uses a REPL; exploration feeds construction. Likewise, prompt engineering is a \emph{technique used within} programmatic prompting---during design spikes, error analysis, and hypothesis generation---but it is not the discipline itself, and organizations that staff for the technique while needing the discipline will fail in the ways Section~I describes.

The empirical record supports the categorical reading. Studies of non-expert prompters show that unassisted interactive iteration is opportunistic, evaluations are anecdotal, and improvements do not accumulate~\cite{ref5}. Liang et al.'s grounded-theory study of twenty professional developers deepens the point: prompt developers reason not from the artifact but from mental models of the model's behavior built through interaction---and even developers who have authored dozens of prompts fail to form reliable such models~\cite{ref22}. If experience does not yield predictive intuition, then intuition cannot be the discipline's correctness mechanism; empirical evaluation must be.

\subsection{Versus Prompt Tuning}
\label{sec:versus-pt}

Prompt tuning (and its relatives prefix tuning and soft prompting) is a \emph{parameter-efficient model adaptation} method: continuous vectors prepended to the input are optimized by gradient descent while model weights stay frozen~\cite{ref15,ref16}. Despite the shared word, it belongs to the machine-learning discipline of model training, not to prompting as considered here: its artifacts are uninterpretable tensors, its method is optimization against a training set, and it requires access to model internals that API consumers typically lack, whereas programmatic prompting operates entirely in the discrete, human-readable input space of a fixed model. Nor does the presence of optimization blur the line: \emph{automatic prompt optimization} (e.g., APE~\cite{ref17}, the optimization loops of DSPy~\cite{ref12}) also optimizes against a metric, but its search space is legible text rather than continuous vectors---which places it, per Section~\ref{sec:definition}, inside the discipline's own practice while prompt tuning remains model adaptation. The two can cooperate---a tuned soft prompt may implement a component whose interface programmatic prompting specifies---but conflating them misleads on every practical question: who does the work, what tools they need, and what can go wrong.

\subsection{Versus LLM Application Development}

LLM application development---building software \emph{on top of} models, not building the models themselves, and in current practice increasingly called \emph{AI agent development}, since even a model wrapped in a system prompt and tool access already constitutes a minimal agent---is the broadest neighbor: everything required to ship an LLM-powered product, from user experience and retrieval infrastructure to orchestration, guardrails, and cost management. It is also the most tempting conflation: if ``AI engineering'' already names the whole, why name a part? Because the part is a coherent sub-discipline. Programmatic prompting is the layer at which system intent is expressed to and enforced upon the model---the behavioral interface of Section~\ref{sec:definition}'s definition, the application's behavioral core---with its own theory, tools, and failure modes, much as database engineering is a sub-discipline within web development despite every web application containing far more than schemas and queries. Naming it allows curricula, tooling, and roles to target it directly instead of dissolving it into full-stack generality.

\subsection{Versus Fine-Tuning}

Full fine-tuning---updating model weights on task-specific data---is not conflated with programmatic prompting but is its principal \emph{alternative implementation strategy}, and the discipline is what makes the choice between them rational rather than ideological: the specification states what either approach must achieve, the evaluation suites measure both on equal footing, and the curated corpus of production traces is the natural source of fine-tuning data, so ``prompted component'' and ``fine-tuned component'' become two implementations behind the same specified interface (Section~\ref{sec:definition}), selected and re-selected on measured cost, quality, and latency. A team that fine-tunes while still carrying the specification and evaluation debt of Section~\ref{sec:background} is guessing twice---about the training data, and about whether the result is better. The boundary between the two can even be stated exactly: treating the model as a stochastic dynamical system, prompting searches for behavior within the fixed model's reachable set, while fine-tuning alters the reachable set itself~\cite{ref30}. The two also compose rather than compete: a fine-tuned model is still invoked through a prompt, and that prompt remains a governed artifact. Fine-tuning changes the substrate; it does not exempt anyone from the discipline.

\section{Related Work}
\label{sec:related}

We have deferred this comparison to the end deliberately: it turns on distinctions---technique versus discipline, descriptive versus prescriptive, tool versus evidence---that only the preceding sections establish. Our argument builds on threads that are now rapidly converging. Reynolds and McDonell first framed prompting as programming~\cite{ref3}; White et al.\ cataloged prompt patterns explicitly by analogy to software design patterns~\cite{ref4}; and Liu et al.\ surveyed prompting as an ML paradigm~\cite{ref20}. The closest descriptive neighbor is Liang et al.'s empirical argument that ``prompts are programs too'' and that their development---``prompt programming''---is a distinct programming phenomenon~\cite{ref22}; their observations (unreliable mental models, interaction-driven development, absent tooling) are the symptoms for which the present paper proposes the treatment, and Villamizar et al.'s research agenda on prompts as software engineering artifacts extends the same empirical line, asking whether systematic prompt management repays its effort~\cite{ref29}---precisely the question our cost-structure hypothesis (Section~\ref{sec:background}) addresses. The closest prescriptive neighbor is Chen et al.'s \emph{promptware engineering}~\cite{ref28}, which likewise treats prompts as first-class software artifacts, diagnoses ad hoc trial-and-error development as a ``promptware crisis,'' and adapts the software lifecycle---requirements, design, implementation, testing, debugging, evolution---to prompts. We differ in what comes first. Promptware engineering supplies a methodology---software engineering's stages transferred onto a new artifact---without first drawing the boundary of the discipline the methodology serves; this paper supplies the boundary: a categorical carving (technique versus discipline) that no stage model expresses, a definition by locus (the behavioral interface) rather than by phases, and an account of prompt debt---its silence and its economics---that explains why the crisis persists, a question the stage-based program does not pose. A terminological confusion also circulates in the other direction: practitioner literature sometimes uses ``programmatic prompting'' itself narrowly, as a synonym for automatic prompt optimization~\cite{ref23}; we claim the term for the discipline, of which machine-optimized wording is one practice among many.

Systems research has supplied artifacts consonant with our argument---LMQL's constrained query language~\cite{ref11}, DSPy's signatures and compilers~\cite{ref12}, BAML's typed prompt-functions with schema-aligned parsing~\cite{ref21}, and chain-authoring environments~\cite{ref9,ref10}---but has framed them as tools rather than as evidence of an emerging discipline. HCI research has documented the failure modes of interactive prompting practice~\cite{ref5} and of informal evaluation~\cite{ref6}. The ML-systems literature diagnosed hidden technical debt~\cite{ref13} and built the MLOps response; we argue prompt-based systems require an analogous but distinct response---distinct because the artifact is legible text and the adaptation loop runs through wording rather than weights. Aljohani and Do empirically catalogue LLM-specific self-admitted technical debt and find prompt debt its largest category~\cite{ref27}; their instrument measures the admitted fraction of the liability, where we argue the silent fraction dominates (Section~\ref{sec:background}).

Finally, maturity models~\cite{ref18} and lifecycle models are the classical instruments by which software engineering converted craft into discipline; we deliberately adopt neither---a restraint that also marks our distance from the stage-based program of promptware engineering~\cite{ref28}---because the field's need, on our argument, is prior to methodology: a boundary for the discipline that any methodology, theirs included, can then serve. What is new here is not any single instrument but the categorical claim: the separation of the interaction technique from the engineering discipline.

\section{Conclusion}
\label{sec:conclusion}

Disciplines are born when a community stops asking ``how do I make this work once?'' and starts asking ``how do we make this keep working?'' Prompting is crossing that line now. This paper has argued that the crossing deserves a discipline---programmatic prompting---a body of practice distinct from the interaction technique of prompt engineering, from the model-adaptation methods of prompt tuning and fine-tuning, and from the encompassing activity of LLM application development. We have proposed a definition resting on the behavioral interface and the five component practices that govern it, and an account of prompt debt whose silent form, we argue, dominates the admitted form that current empirical instruments can measure.

The argument invites two objections worth answering directly. The first: \emph{models will soon be good enough that none of this matters.} Perhaps individual prompts will need less craft; but better models enable more ambitious systems, and system properties---composition, regression, drift, security---are exactly the ones that do not improve with model quality alone. Compilers got better for sixty years; software engineering did not thereby become unnecessary. The second: \emph{this is premature formalization of a fast-moving practice.} We answer that the formalization proposed here is deliberately structural rather than prescriptive of particular tools; specifications, tests, versions, and observability have survived every technology shift since the 1960s, and there is no reason to believe legible-text programs induced from language models will be the exception.

What remains is to equip the discipline, and the anatomy of prompt debt (Section~\ref{sec:background}) says where to dig. The hardest problems sit where specification and evaluation are missing: languages that can state, checkably, what a stochastic linguistic component must do, where constrained generation~\cite{ref11} and schema-aligned parsing~\cite{ref21} today cover only the syntactic layer; behavioral diffing---a principled account of when two prompt versions, or one prompt on two model versions, count as equivalent, with tools that surface the inputs on which they disagree; a compositional semantics for chained model calls; a control-theoretic semantics for the discipline as a whole, building on formalizations of LLMs as discrete stochastic dynamical systems~\cite{ref30}---the specification as reference behavior, the prompt as a synthesized control input, reachability as the feasibility question, and injection as adversarial disturbance; architectural defenses against prompt injection, which remains structurally unsolved so long as instructions and data share one channel~\cite{ref19}; and empirical foundations, beginning with measurement of prompt debt beyond its self-admitted fraction~\cite{ref27} and tests of the adoption-cost hypothesis of Section~\ref{sec:background}. Each deserves fuller treatment than a closing paragraph allows.

Prompt engineering taught us to talk to models. Programmatic prompting is how we will build with them.

\clearpage
\appendices

\section{A Prompt and a Programmatic Prompt}
\label{app:contrast}

The exhibits track the components of Section~\ref{sec:definition}'s definition: the artifact and its contract (A.1), the typed interface (A.2), the locus of contract enforcement (A.3), the evaluation loop and its use across a model migration (A.4), and the machine-closed optimization loop as limit case (A.5). All are drawn from real, current tooling.

\subsection{The Same Task, Two Practices}
The interaction technique, in its natural habitat, is a message typed into a chat window and judged by reading one output:
\begin{quote}
\emph{Summarize this support ticket. Be concise. Don't apologize.}
\end{quote}

The same task as a governed artifact:
\codeinput{ticket-summary.yaml}{a1-ticket-summary.yaml}{yaml}

The template's wording could be identical in both practices; nothing about the words distinguishes them. What distinguishes them is everything around the words---Section~\ref{sec:versus-pe}'s claim in executable form.

\subsection{The Interface, Separated from Its Implementation}
The same component as a DSPy signature~\cite{ref12}---the \emph{what} stated as a typed contract, the \emph{how} (the concrete wording) compiled and replaceable:
\codeinput{a2-dspy.py}{a2-dspy.py}{Python}

The signature survives model migrations and re-optimizations; the induced wording does not have to. This is the interface/implementation split of Section~\ref{sec:definition}, component~(3), as working code---and note that \texttt{compile} demands \texttt{metric} and \texttt{trainset}: the optimizer refuses to run on a team that has not yet paid its specification and evaluation debt.

\subsection{The Output Contract, Enforced at Parse Time}
The same component in BAML~\cite{ref21}, where the return type is a checked contract and schema-aligned parsing repairs the model's reply against it rather than constraining generation:
\codeinput{summarize.baml}{a3-summarize.baml}{BAML}

The alternative locus---constraining generation to the schema at decode time~\cite{ref11}---trades differently, as Section~\ref{sec:definition} discusses; what both share, and what the chat-window prompt of A.1 lacks, is that ``malformed output'' is \emph{defined}.

\subsection{The Loop-Closer: Regression Defined, Then Used}
An evaluation gate in the style of contemporary harnesses, run on every change to the template, the exemplars, or the model pin:
\codeinput{suite.yaml}{a4-suite.yaml}{yaml}

And the report such a gate produces across a substrate change---the artifact held fixed, only the model pin moved:

\begin{table}[!t]
\caption{Evaluation Report Across a Model Migration (G-114)}
\label{tab:migration}
\centering
\begin{tabular}{@{}lcc@{}}
\toprule
\textbf{Metric (G-114)} & \textbf{model-2026-03} & \textbf{model-2026-06} \\
\midrule
Schema validity & 99.8\% & 99.1\% \\
Rubric agreement & 97.4\% & 91.2\% \\
New failure cluster & --- & 9 cases: P1 tickets rated P2 \\
\bottomrule
\end{tabular}
\end{table}

Without the gate, this shift is silent debt (Section~\ref{sec:background}), surfacing on the vendor's schedule; with it, the shift is a diff to triage before rollout---the difference, per Section~\ref{sec:versus-pe}, between an undefined quantity and a poorly behaved one. Drift of exactly this shape has been measured in the field~\cite{ref26}.

\subsection{The Limit Case: A Loop Closed Entirely by Machines}

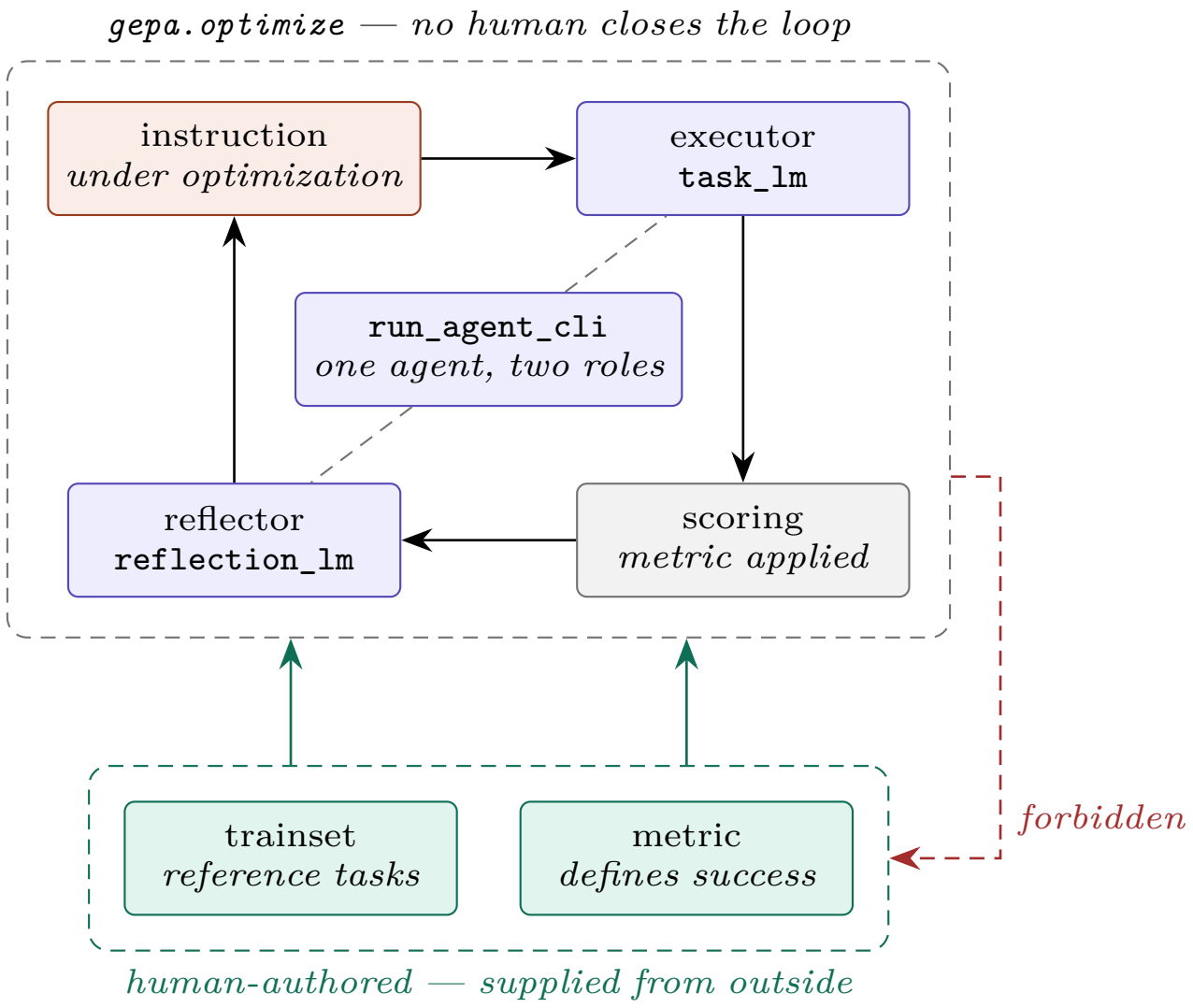
\begin{figure}[t]
\centering
\begin{tikzpicture}[
  font=\scriptsize,
  every node/.style={},
  nd/.style={draw, rounded corners=2pt, align=center, minimum width=2.35cm, minimum height=0.8cm, line width=0.4pt},
  spec/.style={nd, fill=specfill, draw=specborder},
  agent/.style={nd, fill=agentfill, draw=agentborder},
  artifact/.style={nd, fill=artifactfill, draw=artifactborder},
  neutral/.style={nd, fill=black!5, draw=black!55},
  flow/.style={-{Stealth[length=2.2mm]}, line width=0.5pt},
  shared/.style={densely dashed, black!55, line width=0.4pt},
  forbid/.style={-{Stealth[length=2.2mm]}, densely dashed, forbiddenred, line width=0.5pt},
]
\node[artifact] (instr) at (0,0) {instruction\\ \textit{under optimization}};
\node[agent] (exec) at (3.6,0) {executor\\ \texttt{task\_lm}};
\node[neutral] (scor) at (3.6,-2.7) {scoring\\ \textit{metric applied}};
\node[agent] (refl) at (0,-2.7) {reflector\\ \texttt{reflection\_lm}};
\node[agent] (cli) at (1.8,-1.35) {\texttt{run\_agent\_cli}\\ \textit{one agent, two roles}};
\draw[flow] (instr) -- (exec);
\draw[flow] (exec) -- (scor);
\draw[flow] (scor) -- (refl);
\draw[flow] (refl) -- (instr);
\draw[shared] (cli) -- (exec);
\draw[shared] (cli) -- (refl);
\node[draw=black!55, densely dashed, rounded corners=4pt, fit=(instr)(exec)(scor)(refl)(cli), inner sep=8pt,
  label={[font=\scriptsize\itshape]north:{\texttt{gepa.optimize} --- no human closes the loop}}] (loop) {};
\node[spec] (train) at (0.4,-4.95) {trainset\\ \textit{reference tasks}};
\node[spec] (metr) at (3.2,-4.95) {metric\\ \textit{defines success}};
\node[draw=specborder, densely dashed, rounded corners=4pt, fit=(train)(metr), inner sep=7pt,
  label={[font=\scriptsize\itshape, text=specborder]south:human-authored --- supplied from outside}] (humans) {};
\draw[flow, specborder] (train |- humans.north) -- (train |- loop.south);
\draw[flow, specborder] (metr |- humans.north) -- (metr |- loop.south);
\draw[forbid] ($(loop.east)+(0,-0.9)$) -- ++(0.35,0) |- (humans.east)
  node[pos=0.45, right, text=forbiddenred, font=\scriptsize\itshape] {forbidden};
\end{tikzpicture}
\caption{The limit case as a dependency graph. Inside \texttt{gepa.optimize}, a candidate instruction is executed (\texttt{task\_lm}), scored against the metric, and rewritten (\texttt{reflection\_lm}); executor and reflector are the same agent behind \texttt{run\_agent\_cli} (dashed links). The two assets the loop consumes but did not produce---the reference set and the metric---enter from outside. The dashed red arrow marks the step the discipline forbids: the loop authoring its own specification.}
\label{fig:a5loop}
\end{figure}

The following program uses a reflective prompt optimizer~\cite{ref31} to improve an instruction, where both the system executing the instruction and the system rewriting it are the same command-line agent (Fig.~\ref{fig:a5loop}):
\codeinput{a5-gepa.py}{a5-gepa.py}{Python}

Four layers of prompting nest here: the instruction is data under optimization; the optimizer's reflection is itself prompting; the reflector and the executor are one prompted agent; and that agent is a model wrapped in its own system prompt. No human closes the loop. Yet the call cannot execute without two arguments that no model in the loop produced---\texttt{trainset} and the metric it is scored by. The optimizer presupposes, and thereby enforces, exactly the assets of Section~\ref{sec:definition}: this is the reciprocity between optimization and discipline compressed into one function signature. The regress could continue---the optimizer's own reflection prompt could be optimized, the agent could generate its training data---but the final step is the one the discipline forbids, because a loop that authors its own specification optimizes toward its own blind spots. One last detail rewards inspection: \texttt{agent\_task\_lm} concatenates the system instruction and the user input into a single string, collapsing the instruction and data channels whose sharing makes prompt injection structurally unsolved~\cite{ref19}---even the discipline's limit case carries the substrate's oldest vulnerability.

The listing is not hypothetical. We ran it---with a hosted 120-billion-parameter reasoning model (\texttt{gpt-oss-120b}, served via Amazon Bedrock) behind both \texttt{task\_lm} and \texttt{reflection\_lm}---on a support-ticket triage task: an eight-ticket golden set, the metric exact agreement with a human-authored severity rubric. The seed instruction, which specified only the output contract, scored 6/8. After fourteen metric calls the reflector, shown the failing traces, reverse-engineered the specification: it inferred an explicit P1/P2/P3 severity rubric from the golden set's examples and wrote it into the instruction. That candidate scored 7/8; a later mutation regressed to 6/8 and was rejected by the evaluation loop; the budget expired with the rubric-bearing candidate as winner. First, the residual failure is a specification failure, not a capability failure: the one ticket no candidate ever classified correctly sits exactly on the rubric's P2/P3 boundary, an ambiguity only a richer golden set or rubric could resolve---and those are precisely the assets the loop is forbidden from authoring. The machine-closed loop improved only by better approximating the external specification, and its ceiling is that specification's resolution. Second, the winning instruction was itself a defective artifact: the reflector interleaved its chain-of-thought deliberation with the proposed prompt, and the accepted candidate carried that preamble---and a truncation mid-guideline---into every subsequent execution, unnoticed, because nothing in the loop parses the reflector's output. The contract-enforcement requirement of A.3 thus recurses: in a machine-closed loop the new instruction is itself a model output crossing an interface, and it needs the same parse-time validation the discipline demands of any other. A loop that answers to a specification it did not write must also be made to honor contracts on the artifacts it writes for itself.

\end{document}